# THE EMISSION MECHANISM OF GAMMA-RAY BURSTS: IDENTIFICATION VIA OPTICAL-IR SLOPE MEASUREMENTS

Bruce Grossan[1,2], Pawan Kumar[3], George F. Smoot[4,2,5,6,7]

[1] University of California, Berkeley Space Sciences Laboratory, USA
[2] Energetic Cosmos Laboratory, Nazarbayev University, Kazakhstan
[3] University of Texas at Austin, USA
[4] Hong Kong University of Science and Technology, Clear Water Bay, Kowloon, Hong Kong
[5] Université Sorbonne Paris Cité, Laboratoire APC-PCCP, France
[6] University of California, Berkeley, Department of Physics, USA
[7] Lawrence Berkeley National Laboratory, USA

Bruce Grossan: Lawrence Berkeley Lab, 50R-5005, 1 Cyclotron Road Berkeley, CA 94720; Bruce_Grossan@lbl.gov

Pawan Kumar: University of Texas at Austin Department of Astronomy 1 University Station C1400 2511 Speedway - RLM 17.204; pk@surya.as.utexas.edu

George F. Smoot, Lawrence Berkeley Lab, 50-5005, 1 Cyclotron Road Berkeley, CA 94720; GFSmoot@lbl.gov

**Abstract**

There is no consensus on the emission mechanism of $\gamma$-ray bursts (GRBs). A synchrotron model can produce $\gamma$-ray spectra with the empirical Band function form (Band, et al., 1993), from a piece-wise two-power-law electron energy distribution (2EPLS). This synchrotron model predicts that for the *same* $\gamma$-ray spectrum, optical emission can be very different in $f_\nu$ log slope, and in flux relative to $\gamma$-rays, depending on model parameter values. This prediction is consistent with the huge range of optical/$\gamma$ flux ratios observed. The model only allows a small set of $f_\nu$ log slopes in the optical –thereby allowing a clear path to verification or falsification. Measurements of prompt $\gamma$-ray burst emission in the optical thus far give no useful information about the spectral shape within the band, and therefore cannot be used to evaluate such predictions.

We describe an experiment that responds to GRB position alerts with a fast-slewing telescope outfitted with three or more simultaneously recording, high-time resolution cameras, to measure the spectral shape of the prompt optical-IR (OIR) emission. Three channels measure two independent spectral slopes in the OIR region, the minimum information required to evaluate the model, assuming a single dominant component. We propose cross-correlation of $\gamma$ and OIR light curves to verify that a given GRB is single-component dominated, or to model and quantify the contributions from other components. Previous CCD measurements have limited-time resolution due to read noise, limiting cross-correlation analysis. Electron-multiplied CCDS (EMCCDs) can be used to greatly reduce read noise, and allow exposure times of a few hundred ms. Our collaboration has begun a pathfinder experiment, the Nazarbayev University Transient Telescope at Assy-Turgen Astrophysical Observatory (NUTTelA-TAO), with a 70 cm aperture telescope that can point anywhere above the local horizon in $\leq 8$ s, with three simultaneous optical channels. The NUTTelA-TAO is expected to measure the optical slopes of 3-8 GRB/yr, and should provide a clear verification/refutation of the 2EPLS model after a few single-component dominated, sufficiently bright GRBs are detected during prompt emission. A space-based platform would more easily extend the spectral coverage down to near-IR wavelengths, for greater precision in measuring spectral slopes, and increased chance of measuring the self-absorption frequency, which carries valuable information on physical conditions within the GRB jet. Additional science includes detection of dust evaporation due to the UV flash from the burst, which can be used to study dust around a single star at high redshift, independent of host galaxy dust.

# 1. Introduction

### 1.1. Synchrotron GRB Models

The extraordinarily rapid variability of $\gamma$-ray burst (GRB) light curves at E > keV suggested early on that the prompt GRB emission was relativistically beamed. The details of the physical mechanism that produces the emission, however, remain unknown. Previous to ~1997, it was often assumed that this would be Internal Shock Synchrotron (ISS; Rees & Meszaros, 1994; Sari, Narayan, & Piran, 1996; Sari & Piran, 1997, Panaitescu & Mészáros, 1998) with an $f_\nu$ log slope of $-1/2$ (e.g. Piran, 1999) in $\gamma$-ray bands. The expected $-1/2$ $f_\nu$ log slope of ISS, or the predictions of other simple models, are inconsistent with observations, however (e.g. see Preece et al., 1998, regarding the "synchrotron line of death", and Preece et al., 2002). Ghisellini, Celotti, & Lazzati (2000) pointed out theoretical problems with the ISS model, due to the details of cooling time and time scale of measurement, as well as the discrepancy with data.

GRB spectra are now frequently fit in $\gamma$-ray bands with a Band function (Band, et al., 1993), parameterized by a power law with $f_\nu$ log slope $\beta_1$ at E< $E_{peak}$, $E_{peak}$ the energy of the peak in $\nu f_\nu$, and a second power law with $f_\nu$ log slope $\beta_2$ for E>$E_{peak}$. (Additional components have been fit as well; see section 2.2). $\beta_1$ has a broad distribution; *Swift* BAT measured an average of $-0.57 \pm 0.32$ (Sakamoto, et al., 2011). (Likely instrument bias yields different results for different instruments, e.g. Fermi GBM bursts average +0.24, with a width at least as large; Nava, et al., 2011).

Synchrotron GRB models have been covered by a number of authors for both prompt and afterglow emission (e.g., Sari, Piran, & Narayan, 1998; Piran, 1999). Shen & Zhang (2009; hereafter SZ09) used a 2 power law electron population synchrotron emission model (2EPLS) that produces GRB $\gamma$-ray spectra fitting the Band function, to explore the implications for observations in the $\gamma$-ray and optical bands simultaneously. We use their notation and nomenclature for clarity. A piecewise two-power-law electron distribution was used, the first spanning from a minimum energy $E_{min}$ to $E_{peak}$, associated with the peak in $\nu f_\nu$. In the $\gamma$-ray bands, this produces a two power-law emission spectrum

$F_\nu \sim \nu^{\beta_1}$ for $\nu_m < \nu < \nu_{pk}$,   and   (1)

$F_\nu \sim \nu^{\beta_2}$ for $\nu > \nu_{pk}$,   $\nu_m$ corresponding to the minimum electron Lorentz factor, and $\nu_{pk}$ corresponding to the peak of $\nu F_\nu$.

### 1.2. Synchrotron Emission in the IR-Optical bands

The 2EPLS model also produces emission at low frequency optical-IR (OIR) bands; SZ09 described this as four different cases, depending on the location of the self-absorption frequency, $\nu_a$, relative to the optical band ("opt"):

$$F_{\nu,\,opt} \sim \nu^{5/2} \qquad \nu_m < \nu_{opt} < \nu_a,\ \text{Case I}$$
$$F_{\nu,\,opt} \sim \nu^{\beta 1} \qquad \nu_m < \nu_a < \nu_{opt},\ \text{Case II} \qquad (2)$$
$$F_{\nu,\,opt} \sim \nu^{2} \qquad \nu_{opt} < \nu_a < \nu_m,\ \text{Case III}$$
$$F_{\nu,\,opt} \sim \nu^{1/3} \qquad \nu_a < \nu_{opt} < \nu_m,\ \text{Case IV}$$

The spectra, sketched in **Figure 1**, are made up of the multiple power-law segments described above, three segments in Cases I and II, four segments in Cases III and IV, with the lowest frequency break at $\nu_a$. With this 2EPLS model, for different physical parameter sets, the *same γ-ray spectra could have different behaviors at optical-IR (OIR) bands: a finite, but varied, set of spectral shapes (cases I-IV). The value of the $f_\nu$ log slopes in the OIR identifies the case (equation 2). The location of $\nu_a$ relative to the optical has never been measured, therefore, all such cases should be considered.

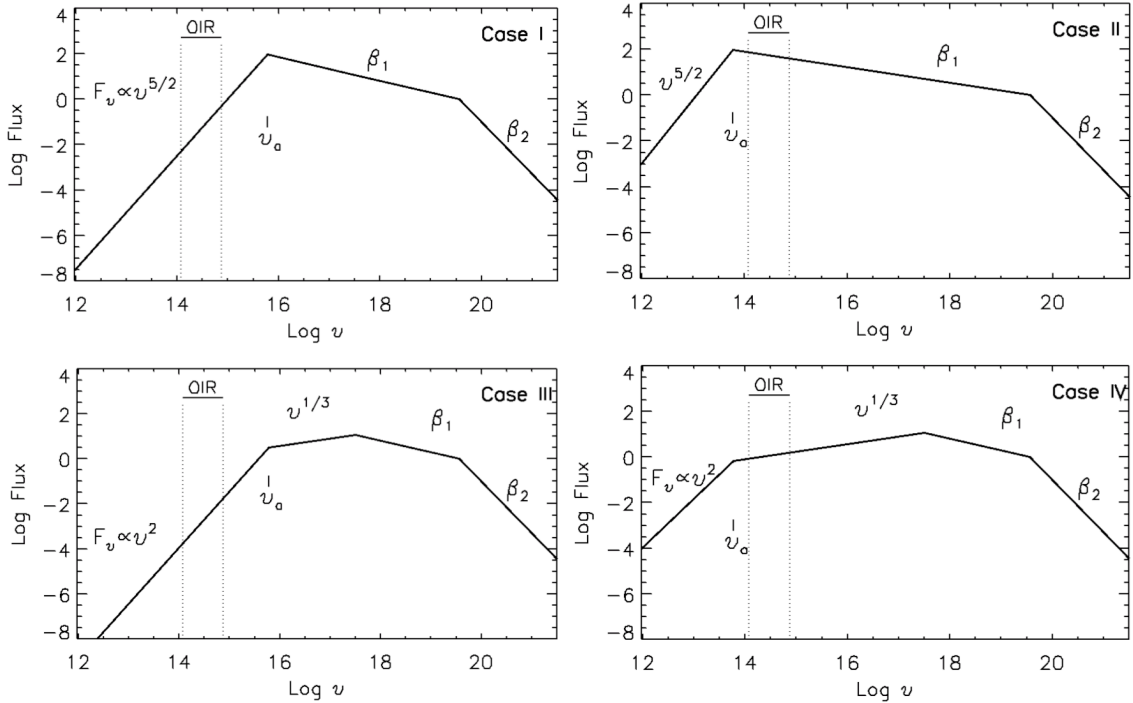

**Figure 1** 2EPLS Spectra. The figures show Cases I-IV from SZ09. In all cases, the EUV- γ-ray spectra are identical, but the spectral shape in the OIR region is different. For each case, the spectral shape (frequency dependence of flux) is given for each segment, along with the location of $\nu_a$, which can be either above or below the OIR bands in these cases.

ROTSE (Akerlof, et al., 2003) and other instruments (including the currently active MASTER-NET; Lipunov, et al., 2010, and in some cases for short times, the SWIFT UVOT; Page, et al., 2019) have detected the optical emission simultaneous with γ-ray emission in ~ dozens of cases. This "prompt" emission is distinct from later "afterglow" emission, which comes from the interaction of the GRB explosion with the surrounding ISM (Rees & Mészáros, 1992, Paczýnski & Rhoads, 1993, Mészáros & Rees, 1993, Mészáros & Rees, 1997). The ratio of optical to $\gamma$-ray emission among the ensemble of GRB is known to range across at least 8 orders of magnitude (080319b had a ratio up to ~$10^4$, whereas several GRB in Rykoff, et al., 2009 had ratios as low as ~ $10^{-5}$), which appears to support an extremely varied set of mechanisms, or suggests no relation between OIR and $\gamma$-ray components. The 2EPLS model, however, naturally gives a wide range of $\gamma$-ray-to-optical flux ratios due to different possible slopes and spectral breaks. Note, however, that only a *restricted set* of OIR log slopes are permitted (equation 2). This is the crucial point that allows the identification of the 2EPLS mechanism. Unfortunately, this means we cannot simply look at the $\gamma$-ray to optical flux ratios measured since the ROTSE era to verify/falsify this mechanism. Virtually all instruments that measured prompt optical emission did so in a single, broad band, usually covering the entire CCD sensitive range, and no reliable measurements of OIR slopes are available. Further progress in this line of inquiry requires a multi-channel instrument capable of measuring the prompt optical-IR spectral shape (POSS). This is the measurement that we explore in this paper.

SZ09 also pointed out that important information on the physical conditions in the jet can de derived from POSS measurements, from the synchrotron self-absorption frequency, $\nu_a$. Depending on information from other sources, it may give the thermal Lorentz factor for the electrons, the B field in the emission region, and the radius of emission – allowing us to map (in the radial dimension) the emission within the jet.

### 1.3. Measuring POSS to Verify the Model

In the rest of this paper we lay out a path forward for verifying or rejecting the 2EPLS mechanism for GRBs. Additionally, through the same observations, we may learn about jet physical conditions via measurement of $\nu_a$. We outline an experimental program to follow-up on GRB alerts with OIR multi-channel protometric imaging at high time resolution. A key part of this project is to control the contamination of other spectral components in the $\gamma$-ray or OIR bands. Using a cross-correlation of the OIR and $\gamma$-ray light curves, we can verify when our OIR measurements are dominated by the same emission process as in the $\gamma$-rays. The multi-channel OIR measurements are used to determine the POSS. Finally, we will test the predictions of the 2EPLS model against the measured slopes, verifying or falsifying this model as the origin of emission.

## 2. Conceptual Experimental Plan

We propose to use fast-pointing OIR telescopes to point to GRB positions provided by $\gamma$-ray instruments, during prompt emission. If such an experiment is not located in space with its own $\gamma$-ray instrument, it is possible to receive real-time position alerts from various $\gamma$-ray space observatories via the Gamma Coordinates Network socket and email alerts (GCN; Barthelmy S. , 2005; Barthelmy, et al., 1995). On receipt of the GRB position, the instrument would point at the GRB and record time-resolved data throughout the prompt emission phase and into the afterglow phase. As GRB spectra are expected to be power-law like continua, we propose that imaging instruments with multiple, simultaneous channels could record light curves and produce a time-history of spectral slopes between each two channels and effectively measure the POSS with much greater simplicity and efficiency than spectrographs.

**Table 1 All SZ09 cases and possible three-channel OIR slopes**

| $\nu_a$ | SZ09 Case | Measured $f_\nu$ Log Slopes | Results |
|---|---|---|---|
| $\nu_a < \nu_1$ | I | N/A | N/A |
| " " | II | $\alpha_{12}=\alpha_{23}=\beta_1$ | Case ID'd, but $\nu_a$ unknown |
| " " | III | N/A | N/A |
| " " | IV | $\alpha_{12}=\alpha_{23}=1/3$ | Case ID'd, but $\nu_a$ unknown |
| $\nu_1 < \nu_a < \nu_2$ | I/II | $\alpha_{12}=f(\nu_a)$; $\alpha_{23}=\beta_1$ | $\alpha_{23}=\beta_1$ => Case I/II ID'd => $\alpha(\nu_1) = 5/2$, => $\nu_a$ |
| " " | III/IV | $\alpha_{12}=f(\nu_a)$; $\alpha_{23}=+1/3$ | $\alpha_{23}=+1/3$ => Case III/VI ID'd; $\alpha(\nu_1) = 2$, => $\nu_a$ |
| $\nu_2 < \nu_a < \nu_3$ | I/II | $\alpha_{12}=5/2$; $\alpha_{23}=f(\nu_a)$ | $\alpha_{12}=5/2$ => Case I/II ID'd; $\alpha(\nu_1) = \beta_1$, => $\nu_a$ |
| " " | III/IV | $\alpha_{12}=2$; $\alpha_{23}=f(\nu_a)$ | $\alpha_{12}=2$ => Case II/III ID'd; $\alpha(\nu_1) = +1/3$, => $\nu_a$ |
| $\nu_3 < \nu_a$ | I | $\alpha_{12}=\alpha_{23}=5/2$ | $\nu_a$ unknown |
| " " | II | N/A | N/A |
| " " | III | $\alpha_{12}=\alpha_{23}=2$ | $\nu_a$ unknown |
| " " | IV | N/A | N/A |

This table gives all possibilities for the measurements of slopes and $\nu_a$ for the 2EPLS synchrotron model, following SZ09, but observed with three filters. The filters are at increasing frequencies $\nu_1$, $\nu_2$, $\nu_3$, giving two log slopes $\alpha_{12}$ and $\alpha_{23}$ between each successive filter pair. $\beta_1$ is the log slope of the low-energy $\gamma$-ray band emission, i.e. at E< $E_{peak}$. We use the abbreviations "=>" for gives or implies, and "ID'd" for identified. In the 4th line below the headings, in the "Measured $f_\nu$ Log Slopes" column, "$\alpha_{12}=f(\nu_a)$" indicates that only $\nu_1$ is on the first power law segment, so $\alpha_{12}$ is not a measure of a power law slope, it varies with $\nu_a$ (see middle panel, Fig. 1). In the Results column, "$\alpha_{23}=\beta_1$ => Case I/II ID'd" should be read, " $\alpha_{23} = \beta_1$ gives an identification of the Case as either I or II". "=>$\alpha(\nu_1) = 5/2$, => $\nu_a$" should be read as, "These cases give the $f_\nu$ log slope at $\nu_1$ to be 5/2, in turn giving the value of $\nu_a$" (i.e. solving for Log $\nu_a$ as the intersection of a line of slope 5/2 through the point (Log $f_1$, Log $\nu_1$) and a line through the two points (Log $f_2$, Log $\nu_2$) and (Log $f_3$, Log $\nu_3$) ).

### 2.1. Synchrotron Cases and Log Slope Space

In sum, all cases of 2EPLS emission allow only the *limited* set of OIR $f_\nu$ log slope values: 1/3, 5/2, 2, or $\beta_1$, the same slope as measured in the $\gamma$-ray band below $\nu_{pk}$ (normally known from the $\gamma$-band observation that initially detects the GRB). This is a well-constrained space of slopes. Under the assumption of a single emission component, and ideal realization of the 2EPLS model, *any other values must reject the* 2EPLS *hypothesis*. In Section 2.2 we discuss validation of the single component condition. In section 7.1 we comment on how non-idealities would affect our measurements.

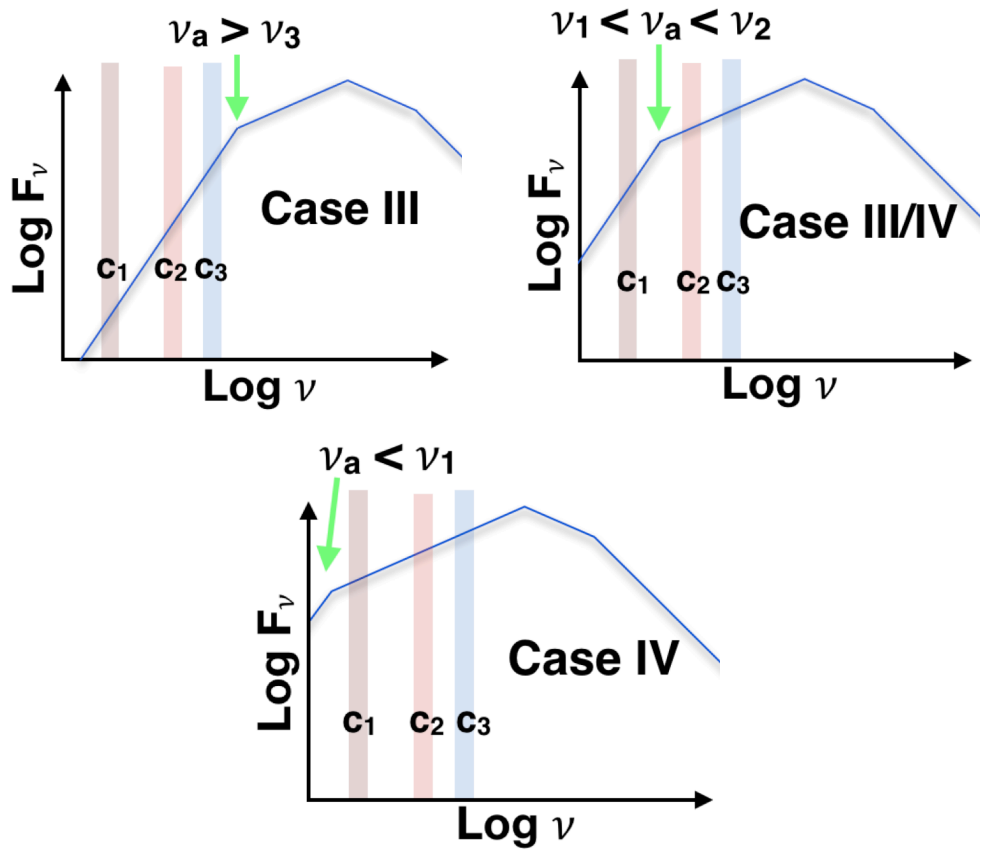

**Figure 2** Selected 2EPLS spectra and measurement channels (filter bands). The figure above illustrates a 4-segment spectrum (cases III and IV; the 3-segment cases are similar). At top left, $\nu_a$ is lower in frequency than any measurement channel (Case IV), so all 3 channels are on the same segment, and the two measured slopes will be equal. At top right, $\nu_a$ is between channels 1 and 2 (in-between cases III and IV). The slope between channels 1 and 2, $\alpha_{12}$, does *not* measure the log slope of any segment because each channel is on a different segment. The slope $\alpha_{23}$ (between channels 2 and 3) does measure the slope of the second-to-the-lowest frequency segment. We then match $\alpha_{23}$, here +1/3, to the allowed OIR log slope values, which uniquely gives us the case, and solve for $\nu_a$. (A similar case occurs when $\nu_a$ is between channels 2 and 3, not shown). At bottom, $\nu_a$ is higher in frequency than any channel (case III), so both slopes are the same (and equal to +2). See Table 1 and Section 2.1. Cases I and II are conceptually the same, with the lowest frequency segment slope=5/2, and the higher frequency segment slope= $\beta_1$.

In the rest of this section we explain how a finite set of measurements of prompt emission can be used to measure the log slope values required to verify or refute the hypothesized spectra. While SZ09 regarded the optical band as if it were a single point on the spectrum, we discuss measurement of spectral slopes by sampling multiple channels (filter bands) in the OIR bands. We assume, for the moment, that the filter width is small enough, and the transition between segments of the spectrum abrupt enough, that there is no significant slope measurement error if $\nu_a$ falls near or within a filter band. We assume that all extinction may be corrected by other information (e.g. from standard dust corrections for the line of sight, and/or observations of the host galaxy).

Any two channels below or above $\nu_a$ would yield a measurement of the log slope of that segment of the spectrum. Two channels are not enough to uniquely identify the case, however, because they could straddle $\nu_a$, in this case, not providing a correct measurement of the slope of any single power law segment.

For three channels at frequencies $\nu_1$, $\nu_2$, $\nu_3$, two log slopes are measured, $\alpha_{12}$ and $\alpha_{23}$. Table 1 enumerates all SZ09 cases, giving the allowed two measured slopes in three OIR channels, and whether or not a determination of $\nu_a$ will result. There are two basic scenarios: all frequencies are on a single power law, or the lower or upper two frequencies could straddle $\nu_a$, illustrated in **Figure 2**. If all channels are above or below $\nu_a$ in frequency, both slopes will be on the same segment of the spectrum, and equal. If the value is one of the 2EPLS allowed slopes, the case is identified (Table 1) and the model is verified, but $\nu_a$ cannot be measured. (As $\nu_a$ is at intersection of two line segments in log space, we must sample two segments to solve for the intersection point.)

If $\nu_a$ is between any two channels, one channel samples one segment on one side of $\nu_a$, and the other two channels give the log slope of the other segment. This log slope can only be one of the allowed slope values, however, allowing us to identify the Case (I-IV) and solve for $\nu_a$. If we do not measure a log slope consistent with an allowed value, we can reject this model.

**Figure 2** illustrates these cases for 2EPLS emission measured in three channels (filters), using the four segment Cases (III and IV). In the first panel, $\nu_a$ is above the frequency of all three channels (Case III), so an $f_\nu$ log slope of +2 must result between any of the channels. In the second panel, the first two channels 1 and 2 straddle $\nu_a$, so $\alpha_{12}$ does not measure the slope of any segment correctly. The $c_2$ and $c_3$ bands would give log slope $\alpha_{23} = +1/3$, identifying this as Case IV for these two cannels, and Case III for channel 1. This identification tells us that channel 1 must be on an $f_\nu$ log slope=2 segment. This in turn allows us to then solve for $\nu_a$ by finding the intersection point of a +2 slope line through (Log $F_1$, Log $\nu_1$) and a line of slope 1/3 through (Log $F_2$, Log $\nu_2$). The problem is the same as $\nu_a$ increases to locations in-between higher frequency filters ($\alpha_{23}=2$, and channel 1 samples the 1/3 slope segment). When $\nu_a$ is below all filter bands (Case IV), then only a single segment is measured again, and we can no longer determine $\nu_a$. All the different scenarios of channel frequency values, $\nu_a$, allowed slopes, and cases for the 2EPLS model are given in Table 1.

2.2. Component Isolation via $\gamma$/OIR light curve correlations

GRB prompt emission is highly variable; therefore light in non-$\gamma$ bands that shows the same time history as in $\gamma$-ray bands must be closely associated with the same physical process. (Specifically, two emission regions in same relativistic flow would have to have separation smaller than $\mathbf{\Delta}$t 2 c $\mathbf{\Gamma}^2$, where $\mathbf{\Delta}$t is the variability time scale and $\mathbf{\Gamma}^2$ is the bulk Lorentz factor of the material.) Many or most GRB are believed to be dominated by a single component (e.g., SZ09), but clearly there is a range of behavior. Optical and $\gamma$ emission *roughly* tracked each other, but not in detail, in 080319B (Racusin, et al., 2008), believed to have more than one emission component. Contrarily, there seemed little correlation between optical and $\gamma$ emission in 990123 (Vestrand, 2005), but only a few points of the optical light curve are available. In the light curve of 110205A, high-energy $\gamma$-ray emission (Suzaku/WAM 110 - 600 keV) and optical/UV emission correlated better than low-energy $\gamma$-rays (*Swift*/BAT 15-150 keV) and optical/UV (Guiriec, et al., 2016). Other bursts, such as 130427A, have been modeled with several components, dominating at different times during the burst (Vestrand, et al., 2014). How can we deal with such complexity? We begin by first identifying the simple single-component dominant bursts, those with the same $\gamma$ and OIR light curve behavior, via cross-correlation. Later, we can build on this understanding, adding linear combinations of light curve models from different mechanisms and locations to understand the more complex ones.

We anticipate that following up GCN alerts to observe prompt emission in the OIR bands from a set of *Swift* GRBs will allow us to build up a "golden sample" of single emission component GRBs. These bursts are then either consistent with synchrotron emission or not, depending on their OIR slopes, and the errors on those slopes. This will be a robust test, given reasonably constrained slope values.

In the case of multi-component bursts, we can make time-dependent multi-component emission models; these will be made under the new constraints of: $\gamma$-ray time series spectra, time-resolved POSS, and multi-band cross-correlations, even within selected epochs of the burst. This should allow the measurement of the amplitude of secondary components, or at least limit their total contribution. Photospheric emission (Mészáros & Rees, 2000; Pe'er, Mészáros, & Rees, 2006) is of particular interest, as adding a thermal component is known to improve the fit in some $\gamma$-ray spectra (Ryde, 2004; Pe'er, Ryde, Wijers, Meszaros, & Rees, 2007; Ryde, et al., 2010; Ryde, et al., 2011; Veres, Zhang, & Mészáros, 2013; Iyyani, et al., 2013), and is expected to have its own time history.

## 3. General Instrumentation Considerations

From the discussion above, an experiment to measure POSS needs to accomplish the following Instrumental Specifications (referred to in the text below as "IS#n", n the specification number):

1. Measure a GRB light curve during most of its prompt emission (we argue this means rapidly pointing to a GRB position alert, i.e. in $\lesssim 10$ s),
2. Measure in multiple channels, such that the POSS is measured, and such that the OIR log slope may be restricted to a small area of parameter space,
3. Measure the light curves in $\gtrsim 10$ time bins, allowing for detailed cross-correlations, and examination of different time periods during the burst.

### 3.1. Rapid-response measurements

A principle challenge for this experiment is IS#1, measurement of the OIR emission simultaneous to $\gamma$-ray emission, a challenge in speed of response. A finite time is required for the BAT instrument to determine an initial position, with a fast time being ~ 10 s (with great variation). We assume no special information comes from the beginning of the GRB, and instead assume that measurements covering the majority of the bright prompt emission are successful. The log T90 (time for 90% of the burst fluence) duration distribution of *Swift* long GRBs peaks at around 70 s (Sakamoto, et al., 2011). Most conventional telescopes require hundreds of seconds to slew across a large fraction of the sky, and the dome motion typically requires even longer. A fast-slewing telescope is therefore required to measure a significant fraction of the prompt emission of most GRBs. An integral field unit spectrometer might do an excellent job (neglecting throughput efficiency penalties) of measuring POSS, but such complex instruments, especially those that would cover the ~ 2'-3' error radius regions of the BAT positions, are much larger and heavier than multi-channel imagers. We therefore propose using a multi-channel imager. Other solutions are indeed possible, e.g. arrays of ultra-wide field cameras covering large fractions of the sky at all times, one array for each color. Such instruments are, thus far, not sensitive (R~10 mag, which can only detect rare bursts), but rapid increases in pixel counts and read speeds may make this an effective instrument in the next generation.

### 3.2. Measurement Band Selection

How do we determine the optimum number and specification of measurement bands for IS#2? If we are to measure $\nu_a$, then the range of our band coverage must include it. For those bursts with detectable optical prompt emission, $\nu_a$ is not likely to be far to the blue of optical; for $\nu_{opt} < \nu_a$, the optical slope must be 2 or 5/2, which is rather steep, so emission would not be bright (i. e. detectable) unless $\nu_a$ were close to the optical or to the red of it. Beyond this rough argument, there are no data to guide us. Therefore, to measure $\nu_a$, the greater the overall spectral coverage, the better. We do not propose measurements at frequencies above optical because most GRBs are at least partially extinguished (Perley, et al., 2009), making the GRBs potentially faint and very hard to detect there. These arguments suggest as many channels as possible from the optical bands to as far in the IR as possible.

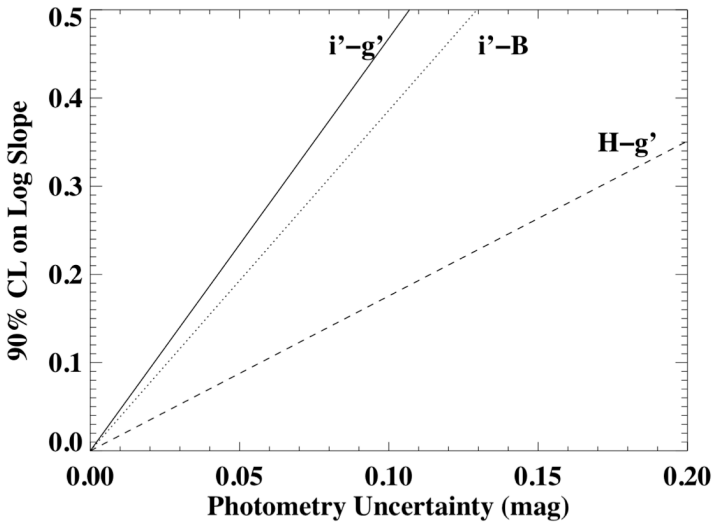

**Figure 3** Slope Confidence Interval vs. Photometric error. The figure shows the 90% CL on the log slope vs. photometric error for various filter pairs, i' and g', i' and B, and g' and H. The inclusion of an H band channel is clearly advantageous.

The primary motivation of our measurement is to measure power law slopes. We need to restrict log slopes to as small a parameter space as possible to verify or reject the 2EPLS model. Such a restriction depends on the photometric accuracy of a measurement, as well as on the separation of the photometric band effective frequency centers (the larger the better). In the simple case of equal photometric error, for various standard filter pairs, we plot the approximate 90% confidence limit on the log slope value (taking the filters to be of infinitesimal width) in Figure 3. Assuming only optical band detectors are available, the photometric accuracy is challenging for short exposures– to obtain a 90% CL of ± 0.25 in log slope, the measurements in both filters, i' and B, must be uncertain by 0.06 mag or less. The measurement is much easier if both optical and NIR (near-IR) channels are available, however, as this makes for a much larger baseline in frequency, requiring only ±0.15 mag precision in H and g' filters for the same confidence interval. As we shall see below, however, the addition of NIR channels adds significant problems on the ground, though it works very well on a space platform.

### 3.3. High Time Resolution via EMCCD

In order to extract detailed information from cross-correlations for IS#3, one needs to move beyond the typical 3- 4 time bins in the prompt optical data (e.g. Vestrand, 2005). Most $\gamma$-ray instruments have high time resolution (e.g., standard BAT GRB light curves, "…bevms.lc", have 64 ms time resolution). Most optical light curves are acquired with CCDs, however, and have poor time resolution (.e.g. ROTSE-III had typically 5 s resolution; Rykoff+09) due to read noise limitations. (Vidicon instruments, e.g. TORTORA (Beskin, et al., 2010), can have higher time resolution, but have the limitation of poor quantum efficiency, and therefore, poor sensitivity). In addition, most astronomical CCDs are optimized for low-noise and have read times of several seconds, which is unacceptable for high time resolution use. Use of an electron-multiplied CCD camera (EMCCD) can essentially solve these problems. With this type of detector, the electron signal is multiplied by $\sim 10^2 - 10^3$ before the read stage of the electronics, essentially boosting all signal above the read noise. Data can be taken at the maximum time resolution of the camera (current commercial EMCCD cameras allow $\sim$ 100 ms), then co-added until the desired SNR is reached, with minimal read noise penalty.

There are two significant downsides to EMCCDs. First, electron multiplication yields loss of dynamic range: as the electron multiplication, or "EM gain" increases, a smaller number of input photons exceeds the maximum number of electrons the system can record (the "saturation level" of the detector). Second, electron multiplication is a stochastic process, and adds an additional factor of $2^{1/2}$ to the noise, the "excess noise" factor (Basden, Haniff, & Mackay, 2003). These effects can be mitigated when operating in a photon counting mode (Daigle & Blais-Ouellette, 2010), if the number of photons is less than 1 per exposure per pixel. In this case, the camera is operated at high EM gain, and a threshold for a binary pixel value is set > 5$\sigma$ above the noise, and only a binary image is recorded. The shot noise of the photon arrivals then dominates, and the read and "excess noise factor" are small, as long as the probability of multiple photons per pixel per frame is small. Adding many images then recovers dynamic range. Successful operation in this mode depends on our sky brightness and other actual measured final performance.

For long GRBs lasting $\sim$ 60 s, if some 40 s of optical data are acquired during the GRB, binned to 300 ms time resolution, that gives 133 simultaneously covered time bins, allowing for detailed cross-correlation studies, even after sub-dividing into a few epochs within the burst.

## 4. The NUTTelA-TAO Ground-Based Pathfinder Experiment

Under a variety of real-world constraints, we set out to quickly start up a minimal experiment to confirm or reject the 2EPLS model. We began with the Section 3 IS#1 for rapid response, identifying a 700 mm aperture commercial telescope that, with suitably modified software, could point and track at any target above our local horizon in less than 8 s. This telescope, the NUTTelA-TAO (Nazarbayev University Transient Telescope at Assy-Turgen Astrophysical Observatory) was installed at the Assy-Turgen Astrophysical Observatory (ATO), through collaboration with the Fesenkov Astrophysical Institute of Almaty, Kazakhstan, and saw first light on Oct. 19, 2018.

To cover as much OIR band as possible, part of Section 3 IS#2, we considered the inclusion of a near-IR (NIR) camera. We could not implement this due to several reasons: First, cost constraints, as a high-quality, NIR-sensitive, HgCdTl detector camera is much more expensive than a CCD or EMCCD camera. Second, the sensitivity of a NIR instrument depends critically on the quality of the site for IR observing. Due to frequent high-moisture conditions, we estimate the site to have poor IR performance (no NIR performance data are available). Additionally, the background at NIR wavelengths (J-Ks) is thousands of times higher than in the optical, at even the best sites. This means that the sensitivity is greatly inferior at these wavelengths. Scaling the actual performance of KPNO telescopes (Massey, et al., 2002; NOAO, 2013) to the particulars of our experiment, our point source sensitivity would be nearly $20^{th}$ mag in 10 s at i'-g'. However, in J-Ks filters, the sensitivity is greater than 3 mag worse for the same KPNO site, and likely still worse for our ATO site. We interpreted this to mean that many bursts would be undetectable with a much less sensitive NIR camera on our small telescope, in up to 10 s exposures. (The exception would be intrinsically bright, but extinguished, bursts.) In addition, our small telescope has weight constraints, and the addition of one NIR camera would require the elimination of one EMCCD camera. Therefore, in the replacement of an EMCCD with a much less sensitive NIR camera, we risk having only two detected channels in many bursts, not enough to definitively verify or reject 2EPLS. For these reasons, we made the decision to reject the proposed NIR camera, at least until a better site was available.

In consideration of Section 3 IS#3, we selected NuVu 1024X1024 EMCCD cameras for the best low-light imaging performance at high frame rate. The weight constraints of our focuser-rotator made it challenging to include more than the minimum 3 channels (Section 2). The resulting burst simultaneous three-channel imager (BSTI) uses two dichroic beam splitters to separate the telescope beam into cameras with g', r', and i' filters. As Sloan filters have little overlap, and we were able to obtain dichroics with sharp reflectance/transmittance transitions, the input to the cameras is very close to the desired Sloan bands. The pixel scale of the telescope gives 0.59 " / pixel, and yields a 10.08' FOV, which is sufficient for BAT's typical 2'-3' position errors. We predict close to 20 mag sensitivity in 10 s in all three bands in photometric conditions, assuming we can get good results in photon counting mode (**Figure 4**). If the actual measured system performance is not appropriate for this mode, the excess noise factor of the EM process will decrease sensitivity by ~ 0.4 mag to ~ 19.6 mag. We anticipate

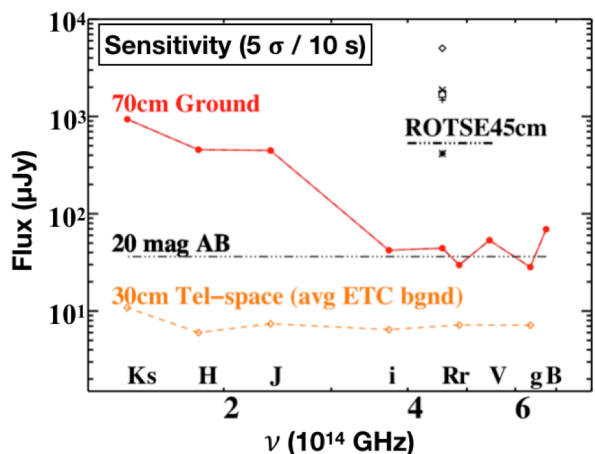

**Figure 4** Sensitivity in 10 s. The 30 cm space telescope is by far the most sensitive. Filter frequencies are as labeled just above the x-axis. The ROTSE average single channel sensitivity is given by the line just below its label; symbols give the prompt detections from (Rykoff, et al., 2009) (but two are off the top edge of the plot). "70 cm ground" gives the estimated performance of a ground-based telescope of 70 cm aperture in good conditions, assuming operation of the EMCCD in photon counting mode. Without photon counting mode, the optical sensitivity only would be worse by 0.38 mag due to the excess noise factor of the EM gain.

commissioning the BSTI in less than a year.

## 5. A Space-Based GRB Mechanism Experiment

A space platform is strongly preferred for this experiment, because of the superior IR performance. As noted above, a fast response favors a low-mass instrument. The need for cryogenic cooling of HgCdTe detectors for ground-based experiments makes such cameras heavy, slows down response, and due to closed-cycle cryogenic refrigerant and cooling hoses, complicates the mechanical design of the instrument. In space, however, it is often possible to use a small amount of Peltier cooling, but mostly passive cooling behind a sunshade, to operate the HgCdTe array at cryogenic temperatures. The NIR camera opto-mechanical design would still typically require a cold stop and associated optics, but this is not a great difficulty or expense. In addition to the hardware simplifications, the huge reduction in background makes a small aperture telescope in space superior to a much larger one on the ground, in the IR. In order to measure log slopes over the largest baseline for the minimum error, and in order to sample the widest band, for the best chance at catching $\nu_a$, we would have at least three channels between H and B (as many as mass and budget permit). Using the ETC average background, we estimate the sensitivity of a 30 cm aperture telescope, the same size as that for UVOT, but with modern HgCdTe and EMCCD detectors, and compare to the sensitivity for a similar 700 mm ground-based system (**Figure 4**). The space platform instrument is about two orders of magnitude more sensitive in the NIR. Due to lower background, we anticipate this system would be operated in photon counting mode, with no excess noise factor.

## 6. Results

### 6.1. Detection Rates for NUTTelA-TAO

The *Swift*-BAT delivers about 87 real-time GCN alerts per year (Grossan, Kumar, Perley, & Smoot, 2014; hereafter GKPS) with 2'-3' error radius. No other real-time alerts are better than 1°-2° per year; for example the systematic error limit of GBM location is greater than 2°, so BAT dominates as the source of sub-degree alert positions. (The SVOM mission Cordier, et al., 2015, should provide localizations on the arcmin scale, and may be launched and begin operations in a relevant time frame, but we conservatively ignore such additional alerts for this section.)

In order for NUTTelA-TAO to detect these alerts, not only must the position be above the local horizon during darkness, with a detectable optical brightness, it must occur without significant clouds in the line of sight, or any other condition that precludes observations (dust, high winds, high moisture, etc.). Simple estimates of these effects tend to grossly overestimate the delivered detection rate. The ROTSE experience provides more practical and useful guidance here. A single ROTSE-III telescope delivered about 3 detections/year, and about 9 total observations during prompt emission (GKPS), claiming an "equivalent" sensitivity (Rykoff, et al., 2009) that scales to about 16.9 in R in 10 s. Assuming the same good weather fraction as for the ROTSE-III site, NUTTelA-TAO must have a prompt detection rate > 3 /yr., as it is significantly more sensitive than ROTSE-III. An estimate to this number, how many more GRB might be detected, is difficult to determine accurately. There are no data on the un-extinguished brightness distribution of prompt optical emission fainter than the ROTSE-III sensitivity. ROTSE-III had no band filter, actually measuring over the CCD sensitivity band. This makes such instruments biased against faint sources that are very red or very blue, i.e. a very red source would yield only noise at the blue end of the response band (and only noise at the red end of the spectrum for a very blue source).

Extinction, giving red observed spectra, is known to be very important. Afterglow measurements of the "P60" sample of GRB, selected only for $\gamma$-ray properties, but observed with sensitive R and near-IR observations, showed that most GRB are modestly extinguished at the source (median $A_V$=0.35 mag); in addition, most afterglows previously thought to have no optical emission were not "dark", but extinguished, with $A_V$ up to a few mag (Perley, et al., 2009). The NUTTelA-TAO's 2.6 mag or greater sensitivity gain over ROTSE-III would detect a number of such moderately extinguished GRB that ROTSE-III would not detect. For example, from the P60 sample, GKPS identified a number of bursts with evidence of extinction that would be detected with sensitive NIR capability. We find that 4 of these have $A_V \sim 0.7 - 1.2$ mag, so they would be too faint for ROTSE-III, but detectable by NUTTelA-TAO. This would give a factor of 1.4 increase in detection rate, for ~4.2 detections/yr.

Now let us consider the maximum gain from NUTTelA-TAO's sensitivity, not just an increased ability to detect the same population as ROTSE-III that have been extinguished. It is certain that some GRB are likely unobservable in the optical: from the P60 sample of 29 GRB, two have Ly$\alpha$ absorption red shifted to the optical, and at least two more have very high extinction, $A_V > 5$ mag. Therefore, not more than 86% are likely optically detectable, and from ROTSE-IIIs 9 total prompt observations/yr., this gives a maximum of 7.8 detections/year. (These numbers would increase if both SVOM and Swift operate at the same time.)

### 6.2. Detection Rates for Space-Based Experiments

A space-based 3+ channel experiment, as described in Section 5, might be co-located with the $\gamma$-ray instrument that produces GRB alerts. In this case, without clouds or bad weather, and without bright daytime sky, only a small exclusion zone around the sun, there are only small limitations on the fraction of well-located GRB that may be measured in the OIR. As in the previous sections, some bursts would be difficult to measure: Ly$\alpha$ absorption red shifted into the OIR would interfere with detections in at least some bands, precluding a full analysis, and some bursts have extremely high extinction, so some 14% of bursts might be lost. The Short GRB population is around 10% of Swift bursts (Sakamoto, et al., 2011), therefore, with no prompt emission known from short bursts, we conservatively exclude all these bursts, giving about 77% of all promptly located bursts. As *Swift*-BAT provided some 87 prompt alerts/yr., our space version experiment, flown on a *Swift*-like platform with a BAT duplicate providing alerts, ~ 75 alerts/yr. would be observed with ~ 100 X the sensitivity of ROTSE-III (**Figure 4**). If we use ROTSE-III's prompt detection rate of ~ 30% (about 3 prompt detections/yr., and about 6 upper limits per year overlapping prompt emission; GKPS), i.e. if we ignore our increased sensitivity, this would give ~ 23 detections/yr. in optical bands. With NIR sensitivity, but requiring we also detect in optical bands to measure slopes, we anticipate a factor of 1.4 increase in detection due to moderately extinguished bursts (Section 6.1), ~ 32 measurements/yr. in all bands. We would anticipate much higher detection rates due to increased sensitivity.

Without a co-located $\gamma$-ray location instrument, detection rates depend on the time fraction for overlap of the FOV of the $\gamma$-ray instrument and the allowed pointing region of the OIR instrument, during their periods of operation. If the spacecraft could be flown within a few degrees of each other in the same orbit, the detection rate would approach that for the co-located case.

## 7. Discussion

We have presented an experiment to measure the slope(s) of prompt GRB emission in the OIR band. These measurements allow verification/ refutation of the 2EPLS model. It also allows us to measure the self-absorption frequency (if present within our bands), and therefore critical parameters of the physical conditions in the jet. We also showed that a small space-based experiment would be a significant improvement in performance over an earth-based instrument, because of superior NIR performance. (We note that, unfortunately, no proposed or planned published space missions have prompt pointing *and* multi-channel OIR measurement capabilities required for this experiment, only one or the other.) For any implementation of this experiment, detection of optical emission from a single-component dominant burst (verified with correlated optical and $\gamma$ emission) would give a straightforward evaluation of the 2EPLS mechanism for that burst. We anticipate a real experiment would, at the minimum, yield a clear evaluation for a "golden" sub-sample of such single-component bursts. If the ensemble of measured log slope values tend to cluster around the allowed 2EPLS log slope values (the expected spread of the measured values around the allowed values due to measurement error), there could be little doubt that this model is important in many of these cases.

We note that there is a known instrumental bias that affects the measurement of $\beta_1$ by $\gamma$-ray band instruments, particularly in instruments that do not cover $E_{peak}$ of a burst, such as is often the case with BAT. In joint fits of GBM and BAT vs. fits with BAT only BAT (Virgili, Qin, Zhang, & Liang, 2012), it can be seen that the joint fits have a log slope lower by $\sim 0.3$ than that for BAT. As explained by these authors, this is due to the curvature of the spectrum near $E_{peak}$. Our data analysis should then include a correction for BAT slopes via comparison of the subset of bursts measured simultaneously with higher energy instruments and the resulting $\beta_1$ values with and without the additional high energy data. A conservative overall approach might first look at OIR log slope values $\alpha_i$ in the range of the static allowed values (1/3, 5/2, 2), which are significantly different from the average corrected $\beta_1$ value $\sim -0.9$, to look for clustering about the static values. In a second step, a histogram of $\alpha_i - \beta_{1i,corrected}$ should also show clustering around zero in the 2EPLS case, again within measurement (and correction) uncertainties. In the overall picture, the instrumental bias in measuring $\beta_1$ would cause a significant problem for the basic concepts here only if Case II were a significant fraction of the bursts, *and* the correction increased the spread of $\alpha_i - \beta_{1i,corrected}$ significantly compared to the measurement uncertainty.

### 7.1. Non-ideal deviations from the 2EPLS, and other models

It may be the case that there is a transition region between the power law segments of the predicted 2EPLS emission spectrum, such that in some significant frequency interval around the intersection of two segments, a slope between that of the two adjacent segments would be measured. *Infinitely* sharp transitions between two segments are unrealistic at some level, resulting from unrealistic theoretical assumptions, e.g. that an electron population has a uniform power-law distribution between two energies, and zero electrons above or below those energies.

A transition region covering a large frequency interval could complicate the application of our method, but would not invalidate it. First, consider that $\nu_a$ could be, in most bursts, decades in frequency away from our measurements (there is no measurement suggesting otherwise); even a large transition frequency interval would then have minimal effect on our measurements. Second, consider an experiment that ran until e.g. ~ dozens of single-component bursts were measured. If only a few bursts showed the allowed 2EPLS log slopes (i.e. had sharper transitions and/or $\nu_a$ far from the measurements), we could say we understand that subset, and make progress further modeling and understanding these alone. (The remainder would be said to be either a different mechanism, or possibly have a large transition region.) Thirdly, we note that there is sufficient information in our measurements to indicate this transition is present, rather than a completely different mechanism. Consider the case with $\nu_a$ greater or less than any measurement frequency. Here, one of the two slopes would be significantly deviating from an allowed value, the other would be closer to this allowed slope, as it is farther away from the nominal segment intersection. The ensemble of these measurements would tend to cluster near the allowed value, but asymmetrically, unlike a random error or random slope value. Next, consider the cases where $\nu_a$ was inferred to be between two channels. Unless the transition region were ~ half the channel spacing or larger, one of the slopes would be consistently close to an allowed slope, the other, in-between allowed values. This pattern of behavior would be an indication that a transition parameter would have to be fit to the models. Finally, if the data showed little preference for allowed log slopes, this would mean, quantitatively, that either the 2EPLS is not the correct mechanism, or the transition width parameter was large compared to the size of the individual segments. In the latter case, there would be little practical difference between a non- 2EPLS mechanism, and 2EPLS with a very large transition band; the physics that would make a very large transition region would essentially be a different model anyway, with either significant deviations from the 2EPLS model or other processes dominating the measured spectrum in the OIR.

Finally, we emphasize that the 3 (or more) channel, high time-resolution data would be a very useful *general* tool for evaluating any type of model with any OIR emission. Observing a large sample of GRB, and building up a data base of POSS measurements, should permit understanding of the diversity of the population of bursts and mechanisms.

### 7.2. Additional Science from POSS Measurements

**Studying Short-Type GRBs -** In the bulk of this paper we ignored Short-type GRBs (SGRBs), as it is not possible to measure the prompt emission during the < 2 s duration of these events with our experiment. The GCN localization time plus slew time is larger than 2 s. However, 25% to 40% of SGRB include "extended emission" (EE), which can last more than $10^2$ s, is softer than the initial pulse, or SGRBs without EE, and may be more like that in Long GRBs (Norris & Bonnell, 2006, Norris & Gehrels, 2008, Norris, Gehrels, & Scargle, 2011). Studies of these bursts will therefore be valuable for understanding the mechanisms of SGRBs, EE, and the common properties of all bursts. As SGRBs are now a known target of gravitational wave telescopes (Abbott, et al., 2017), enormous amounts of data from massive observing campaigns will likely accompany these events; the addition of prompt optical information would be of great interest.

**Dynamic Dust Measurements In Individual, High-z Bursting Star Systems-** Another phenomenon that may be studied with our instrumentation is very-rapid destruction of circumburst dust by prompt optical-UV GRB emission (e.g., Waxman and Draine 2000; Perna et al. 2003). If this occurs, the source would rapidly brighten and the color would change from very red to blue as the radiation vaporizes the dust, on the time scale ~ 60 s. By the end of the prompt phase, local dust should be destroyed and no longer cause extinction, allowing estimation of absorption at earlier times. Direct detection of this process would open new avenues for studying GRB environments and progenitors; in particular, only the dust local to the GRB would be destroyed and change extinction properties, allowing separation of local and host galaxy dust effects. This process gives perhaps the only tool to study dust in individual star systems independent of host dust, and because of the brightness of GRBs, it could be used to extraordinary red shifts. This measurement requires the rapid-response and multi-channel measurements of prompt emission unique to this experiment. This topic is discussed in (GKPS), for a two-channel instrument; with three channels, a better separation of intrinsic slope and extinction could be made, and under ideal conditions, the curvature of the dust law could be checked for additional information on the composition of the dust. (Perna, Lazzati, & Fiore, 2003; Waxman & Draine, 2000)

## Acknowledgements

The authors wish to thank the management and staff of the Assy-Turgen Astrophysical Observatory, Kazakhstan and the Fesenkov Institute, Almaty, Kazakhstan, especially Maxim Krugov, for invaluable help with the NUTTelA-TAO enclosure and systems, logistics of observation and instrument maintenance, and a warm welcome to the Kazkhstan Astronomy Community. We also wish to thank the staff of the Energetic Cosmos Laboratory, Nazarbayev University, Kazakhstan, for support of the authors during visits and laboratory work. This work received support from the Energetic Cosmos Laboratory.

# REFERENCES

Abbott, B., Abbott, R., Abbott, T., Acernese, F., Ackley, K., Adams, C., et al. (2017). *PhysRevLett*, *119*, 161101.

Akerlof, C., Kehoe, R., McKay, T., Rykoff, E., Smith, D., Casperson, D., et al. (2003). The ROTSE-III Robotic Telescope System. *PASP*, *115*, 132.

Band, D., Matteson, J., Ford, L., Schaefer, B., Palmer, D., Teegarden, B., et al. (1993). *413*, 281.

Barthelmy, S. (2005). Retrieved May 16, 2019, from GCN: The Gamma-ray Coordinates Network (TAN: Transient Astronomy Network): https://gcn.gsfc.nasa.gov/gcn/

Barthelmy, S. B., Cline, T., Gehrels, N., Fishman, G., C. Kouveliotou, & C. A. Meegan. (1995). *A&SS*, *231*, 235.

Basden, A. G., Haniff, C. A., & Mackay, C. D. (2003). *MNRAS*, *345*, 985.

Beskin, G., Bondar, S., Karpov, S., Plokhotnichenko, V., Guarnieri, A., Bartolini, C., et al. (2010). From TORTORA to MegaTORTORA—Results and Prospects of Search for Fast Optical Transients. *AIA, 2010*, 53.

Cordier, B., Wei, J., Atteia, J. -L., Basa, S., Claret, A., Daigne, F., et al. (2015). *Proceedings of the conference Swift: 10 Years of Discovery*, (p. 5).

Daigle, O., & Blais-Ouellette, S. (2010). *Proc. SPIE, 7536*, 753606-1.

Ghisellini, G., Celotti, A., & Lazzati, D. (2000). Constraints on the emission mechanisms of gamma-ray bursts. *MNRAS, 313*, 1.

Grossan, B., Kumar, P., Perley, D., & Smoot, G. F. (2014). *PASP, 126*, 885.

Guiriec, S., Kouveliotou, C., Hartmann, D., Granot, J., Asano, K., Mésáros, P., et al. (2016). A Unified Model for GRB Prompt Emission from Optical to $\gamma$-rays; Exploring GRBs as Standard Candles. *ApJL, 831*, L8.

Iyyani, S., Ryde, F., Axelsson, M., Burgess, J., Guiriec, S., Larsson, J., et al. (2013 ). *MNRAS, 433*, 2739.

Lipunov, V., Kornilov, V., Gorbovskoy, E., Shatskij, N., Kuvshinov, D., Tyurina, N., et al. (2010). *AdvAst, 2010*, 30L.

Massey, P., Armandroff, T., De Veny, J., Claver, C., Harmer, C., Jacoby, G., et al. (2002). *KPNO Direct Imaging Manual.* KPNO.

Mészáros, P., & Rees, M. (1993). *APJ, 405*, 278.

Mészáros, P., & Rees, M. (1997). *ApJl, 476*, 232.

Mészáros, P., & Rees, M. (2000). *ApJ, 530*, 292.

Nava, L., Ghirlanda, G., Ghisellini, G., & Celotti, A. (2011). *MNRAS, 415*, 3153.

NOAO. (2013, Sept.). *WHIRC User Information.* Retrieved May 15, 2019, from WHIRC Home Page: https://www.noao.edu/kpno/manuals/whirc/whirc.user.html

Norris, J. P., Gehrels, N., & Scargle, J. D. (2011). *735*, 23.

Norris, J., & Bonnell, J. (2006). *ApJ, 643*, 266.

Norris, J., & Gehrels, N. (2008). *AIP, 1133*, 112.

Paciesas, W., Meegan, C., von Kienlin, A., Bhat, P., Bissaldi, E., Briggs, M., et al. (2012). *199*, 18.

Paczýnski, B., & Rhoads, J. (1993). *ApJl, 418*, 5.

Page, M. J., Oates, S. R., De Pasquale, M., Breeveld, A. A., Emery, S. W., Kuin, N. P., et al. (2019). *MNRAS, 1845*, 1793.

Panaitescu, A., & Mészáros, P. (1998). *ApJ, 501*, 772.

Pe'er, A., Ryde, F., Wijers, R., Meszaros, P., & Rees, M. (2007). A New Method of Determining the Initial Size and Lorentz Factor of Gamma-Ray Burst Fireballs Using a Thermal Emission Component. *ApJL, 664*, 1.

Perley, D., Cenko, S., Bloom, J., Chen, H.-W., Butler, N., Kocevski, D., et al. (2009). The Host Galaxies of Swift Dark Gamma-ray Bursts: Observational Constraints on Highly Obscured and Very High Redshift GRBs. *AJ, 138*, 1690.

Perna, R., Lazzati, D., & Fiore, F. (2003). Time-dependent Photoionization in a Dusty Medium. II. Evolution of Dust Distributions and Optical Opacities. *ApJ , 585*, 775.

Piran, T. (1999). Gamma-ray bursts and the fireball model. *PhR , 314*, 575.

Preece, R. D., Briggs, M. S., Mallozzi, R. S., Pendleton, G., Paciesas, W., & Band, D. (1998). *ApJL , 506*, 23.

Preece, R., Briggs, M., Giblin, T., Mallozzi, R., Pendleton, G., Paciesas, W., et al. (2002). *ApJ , 581*, 1248.

Racusin, J., Karpov, S., Sokolowski, M., Granot, J., Wu, X., Pal'Shin, V., et al. (2008). Broadband observations of the naked-eye gamma-ray burst GRB 080319B. *Nature , 455*, 183.

Rees, M. J., & Mészáros, P. (1992). *MNRAS , 258*, 41.

Rees, M. J., & Meszaros, P. (1994). *ApJL , 430*, 93.

Ryde, F. (2004). The Cooling Behavior of Thermal Pulses in Gamma-Ray Bursts. *ApJ , 614*, 827.

Ryde, F., Axelsson, M., Zhang, B., McGlynn, S., Pe'er, A., Lundman, C., et al. (2010). Identification and Properties of the Photospheric Emission in GRB090902B. *ApJL , 709*, 172.

Ryde, F., Pe'Er, A., Nymark, T., Axelsson, M., Moretti, E., Lundman, C., et al. (2011). Observational evidence of dissipative photospheres in gamma-ray bursts. *MNRAS , 415*, 3693.

Rykoff, E., Aharonian, F., Akerlof, C., Ashley, M., Barthelmy, S., Flewelling, H., et al. (2009). Looking Into The Fireball: ROTSE-III And Swift Observations Of Early Gamma-Ray Burst Afterglows. *ApJ , 702*, 489.

Sakamoto, T., Barthelmy, S., Baumgartner, W., Cummings, J., Fenimore, E., Gehrels, N., et al. (2011). The Second Swift Burst Alert Telescope Gamma-Ray Burst Catalog. *ApJS , 195*, 2.

Sari, R., & Piran, T. (1997). Cosmological gamma-ray bursts: internal versus external shocks. *MNRAS , 287*, 110.

Sari, R., Narayan, R., & Piran, T. (1996). *ApJ , 473*, 204.

Sari, R., Piran, T., & Narayan, R. (1998). *ApJL , 497*, 17.

Shen, R.-F., & Zhang, B. (2009). Prompt optical emission and synchrotron self-absorption constraints on emission site of GRBs. *MNRAS , 398*, 1936.

Veres, P., Zhang, B.-B., & Mészáros, P. (2013). Magnetically and Baryonically Dominated Photospheric Gamma-Ray Burst Model Fits to Fermi-LAT Observations. *ApJ , 764*, 94.

Vestrand, W., Wozniak, P., Wren, J., Fenimore, E., Sakamoto, T., White, R., et al. (2005). A link between prompt optical and prompt γ-ray emission in γ-ray bursts. *Nature , 435*, 178.

Vestrand, W., Wren, J., Panaitescu, A., Wozniak, P., Davis, H., Palmer, D., et al. (2014). The Bright Optical flash and Afterglow from the Gamma-Ray Burst GRB 130427A. *Science , 343*, 38.

Virgili, F., Qin, Y., Zhang, B., & Liang, E. (2012). Spectral and temporal analysis of the joint Swift/BAT-Fermi/GBM GRB sample. *MNRAS , 424*, 2821.

Waxman, E., & Draine, B. (2000). Dust Sublimation by Gamma-ray Bursts and Its Implications. *ApJ , 537*, 796.